# Comparative Analysis of Hybrid Precoding Optimization Approaches for Millimeter Wave Massive MIMO System

Om Nath Acharya[1], Ram Kaji Budhathoki[1], Santosh Shaha[1]

[1]Kathmandu University, Department of Electrical and Electronics Engineering, School of Engineering,
Dhulikhel, Nepal
*acharya.om@ku.edu.np*, *ram.budhathoki@ku.edu.np*, *santosh96shaha@gmail.com*

**Abstract:** *The millimeter wave massive MIMO wireless communication is an emerging technology for next generation communication. This technology implements the hybrid analog and digital beamforming structure which helps to maximize the spectral efficiency, minimize the system complexity and achieve high diversity gain. The conventional digital beamforming technique needs one radio frequency (RF) chain per antenna element. High power consumption, significantly high cost of RF chain components per antenna and complex signal processing task at base band makes digital beamforming unsuitable for implementation in massive MIMO system. Hybrid beamforming takes the benefits of both analog and digital beamforming schemes. Near optimal performance can be achieved using very few RF chains in hybrid beamforming method. By optimizing the hybrid transmit and receive beamformers, the spectral efficiency of a system can be maximized. The hybrid beamforming optimization problem is non-convex optimization problem due to the non-convexity of the constraints. By using the traditional techniques, it is very difficult to obtain the optimal solution over the given precoding optimization parameters. In this regard, deep neural network based approach is implemented to solve the optimization problem. The auto encoder driven neural network learning goes through the numerous training iterations with large amount of samples and hence it is able to know the complex characteristics of a wireless channels. In this research work, the millimeter wave massive MIMO system performance implementing the hybrid precoding based on the conventional methods and deep neural network is studied considering the spectral efficiency, bit error rate and complexity. It is shown that the deep neural network based approach for hybrid precoding optimization outperforms conventional techniques by solving the non-convex optimization problem. Moreover, the DNN model accuracy is also analyzed by observing the training accuracy and test accuracy.*

## 1. Introduction

New standards in wireless communications are being proposed as the number of users is increasing rapidly and better quality of service is expected. The service providers try to transport low latency high quality videos and other applications for wireless devices. For this purpose, millimeter wave massive MIMO [1] [26] is considered to be the key technology for the enhancement of system performance. Massive MIMO system is a part of wireless technology where base stations and user equipment's are equipped with large number of antennas to increase the spectral efficiency and the energy efficiency. Massive MIMO helps to increase system capacity by spatial multiplexing and also increases link reliability by reducing bit error rate during transmission. Massive MIMO system mitigates the interference from the perspective of beamforming [2]. By using millimeter wave frequencies, it is possible to reduce the size of the antenna which in turn makes possible for the implementation of massive MIMO system. The millimeter wave falls to the radio spectrum 30 GHz to 300 GHz that corresponds to the wavelength between one and ten millimeters [3], [4]. The main advantages of millimeter wave propagation are the availability of large amount of spectral width. Due to the availability of large bandwidth, millimeter wave wireless link can achieve very high capacity. Millimeter wave links have narrow beams for equivalent antenna system. Generally, millimeter wave (mmWave) systems need high base station density for the acceptable coverage because of the peak power limitation and extreme path loss. The very dense mmWave networks have to consider the interference issue. More users connected in a network increase the chance of interference. Special characteristics of mmWave communication such as high path loss, peak power limitation and signal blockage have added system design complexity [5] - [8].

Beam forming is the suitable technique for providing the solution for such problems. At millimeter wave frequencies, the physical size of antenna is very small which in turn helps for the hardware implementation and large scale beam forming. Beamforming, also known as spatial filtering is a crucial technique in wireless communications. The aim of beamforming is to generate beam of a necessary shape and direct that generated beam in a desired direction. There are three types of beamforming techniques namely analog beamforming, digital beamforming and the hybrid beamforming but the hybrid beamforming is considered to be very effective beamforming technique and latest researches are focused on hybrid beamforming architecture. The analog beamforming architecture is based on the phase shifters. So, the optimization problem becomes very complex due to the constant modulus constraint. For full digital beamforming techniques, a dedicated radio frequency (RF) chain is required for each antenna element which in turn degrades the system performance in terms of cost, high power consumption, high computational complexity and hardware complexity. To overcome such problems, hybrid beamforming is introduced. Hybrid beamforming technique is considered for 5G and beyond cellular communication network. Hybrid beamforming implements a low dimensional digital precoder that follows high dimensional analog precoder and generates the pencil beams. The conventional algorithms such as orthogonal matching pursuit [2] manifold optimization [9], MMSE [3] are proposed for finding the hybrid beamforming precoder and combiner. But these algorithms are time consuming and require perfect channel state information [8]- [12].

The deep learning based method introduced for the hybrid beamforming provides the efficient solution for the complex nonlinear optimization problem. The deep neural network (DNN) is trained to obtain the optimized beamformer which helps to maximize the spectral efficiency [13]. Even in the channel changing environment, deep learning provides the solution faster and hence it is considered to be the efficient techniques for reducing the latency in the cellular networks. Deep learning goes through the numerous training iterations with large amount of samples and hence it is able to know the complex characteristics of a wireless channels [14].

In different literatures, hybrid structure for the beamforming has been studied. Reference [12] gives a comprehensive survey of the different hybrid structures. The fully-connected and partially-connected structures, were proposed to design a hybrid analog and digital precoding algorithm which can minimize the cost of RF chains. Reference [15] demonstrated a low-complexity alternating minimization precoder by considering an orthogonal constraint on the digital precoder. Ayach et al.in reference [2] has shown receive baseband combiners with the aim of minimizing

MSE between transmitted and received signals. In [9], two precoders based on the principle of manifold optimization and particle swarm optimization were proposed. An energy-efficient hybrid precoding for partially-connected architecture was proposed in [16]. Reference [17] presented achievable rates of hybrid precoding in multi-user multiple input multiple-output (MU-MIMO) system when employing only one RF chain per user and has shown the effect of phase error on hybrid structure performance. Also, [18] proposed an iterative hybrid beamforming algorithm for the single user in mmWave channel that can approach the unconstrained digital beamforming solutions. In reference [8], geometric mean decomposition (GMD) based scheme is adopted, but this approach cannot properly address the non-convexity of constraint in analog part.

Now a day, deep neural network (DNN) based methods are being used to solve the complex non-convex optimization problems. In [19], deep learning assisted millimeter wave massive MIMO framework is proposed for the efficient hybrid precoding where selection of the hybrid precoder takes place through the training of DNN and the optimized values for the precoder are considered for maximizing the spectral efficiency. In [20], unconstrained optimum beamformers at the transmitter side and the receiver side is found by using DNN architectures for learning the singular value decomposition. For the enhancement of the performance in deep learning aided hybrid beamforming's, there are still issues to be addressed.

## 2. Hybrid Beamforming Modeling for Massive MIMO

2.1 System Model

The hybrid beamformer consists of a digital baseband precoder and the analog precoder which are represented by $F_D$ and $F_A$ respectively. The $N_t$ and $N_r$ represent the respective number of antennas implemented at the transmitter and the receiver side. In this hybrid system, the transmitter uses $N_{RF}^t$ RF chains and receiver applies $N_{RF}^r$ RF chains as shown in figure 1.

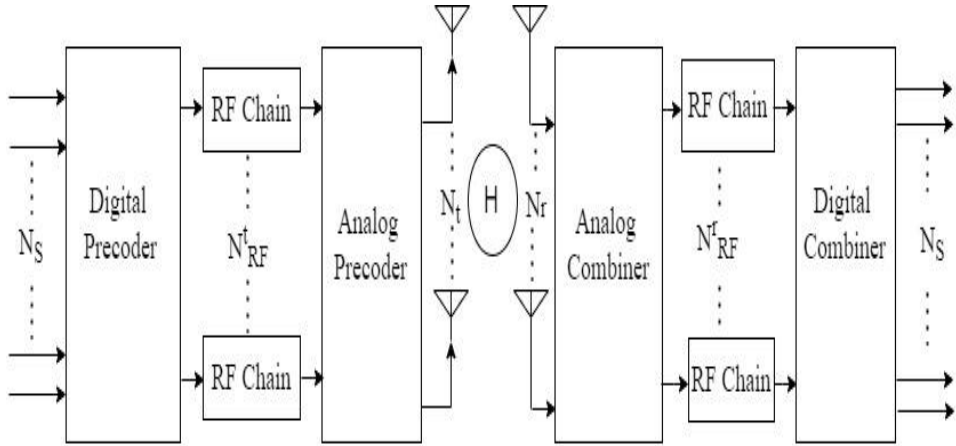

**Figure 1:** Hybrid Beamforming System Model

For $N_s$ to be the number of data streams, the size of digital precoder, $F_D$ is $N_{RF}^t \times N_s$ and the $N_t \times N_{RF}^t$ is the size of the analog precoder, $F_A$. For multiple data transmission, the constraints must be considered is: $N_s \leq N_{RF}^t \leq N_t$. The baseband signal transmission vector, $S$ is given by $S = [s_1, s_2, \ldots s_{\{N_s\}}]^T$. The discrete time transmitted signal is written as $X = F_D F_A S$. The dimension of the S is $N_s \times 1$ which satisfies the condition: $E[SS^H] = \frac{1}{N_s} I N_s$ and $X = [x_1, x_2, \ldots x_{N_t}]^T$. The system constraint is given as: $|F_D F_A|_F^2 = N_s$. Now, the received signal [2] can be expressed as:

$$Y = \sqrt{\rho} HX + N \qquad (1)$$

or

$$Y = \sqrt{\rho} H F_D F_A S + N \qquad (2)$$

where H is $N_t x N_r$ channel matrix, N is additive noise vector and ρ is average received power. $F = F_D F_A$ is known as hybrid precoding matrix of size $N_t X N_r$. Let m be the number of antennas in a sub connected structure where $m = \frac{N_t}{N_{RF}^t}$ and the number of phase shifter required for phase adjustment in analog part is also m.

The hybrid beamforming architecture is classified as fully connected and sub-connected configurations. In fully connected configuration, each RF chain is connected to corresponding antenna which is not true in case of partially connected architecture. In fully connected structure, of analog beamformer, all the receive antennas are connected to each RF chain through the phase shifter network as shown in the figure 2.

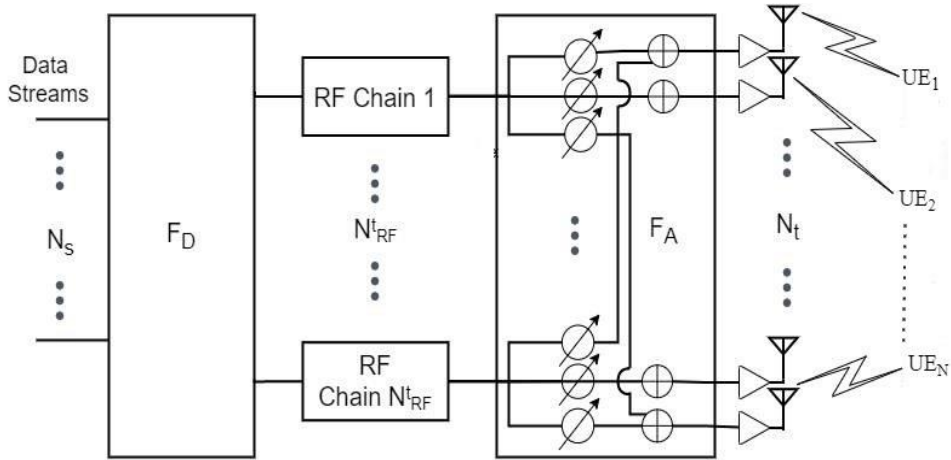

**Figure 2:** Fully Connected Hybrid beamforming System Model

Although the fully-connected architecture has the potential of achieving the full beamforming gain for each RF chain, it requires complex circuitry and consumes relatively high power [12]. In partially connected analog beamforming, a subset of antennas is connected to each RF chain as shown in figure 3. For sub-connected structure, the digital precoding matrix is a diagonal matrix and analog precoding matrix is block diagonal matrix. The analog precoder $F_A$ is represented as:

$$F_A = \begin{bmatrix} a_1 & 0 & \ldots & 0 \\ 0 & a_1 & \ldots & . \\ \vdots & \vdots & \vdots & . \\ 0 & 0 & . & a_k \end{bmatrix} \qquad (3)$$

where $a_k$ is the analog beamforming vector in the $k_{th}$ subarray. The received signal obtained in equation 1 is further subjected to RF combiner and baseband combiner. Let $W_A$ and $W_D$ be the RF combiner and the baseband combiner respectively. The dimensions of $W_A$ and $W_D$ are $N_r \times N_{RF}^r$ and $N_{RF}^r \times N_s$ respectively. The received signal [8] - [12] after the application of the RF combiner and baseband combiner is given by:

$$\bar{Y} = \sqrt{\rho}\, H\, W_A{}^H\, W_D{}^H\, F_D F_A S + W_A{}^H\, W_D{}^H\, N \qquad (4)$$

where $\bar{Y} = [\bar{y}_1, \bar{y}_1, \ldots, \bar{y}_{N_s}]^T$ and $W_A{}^H$ represents the Hermitian transpose of $W_A$.

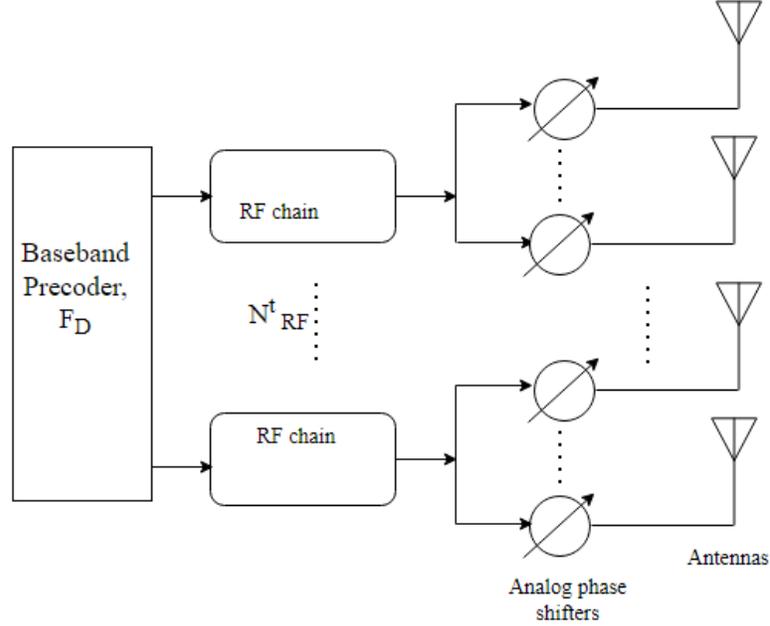

**Figure 3:** Partially Connected Hybrid Beamforming System Model

## 2.2 Channel Model

The traditional channel model is not applicable for millimeter wave massive MIMO system because of the fact that millimeter waves have different propagation characteristics. From the literature [11], the Saleh-Valenzuela model is considered for millimeter wave massive MIMO system and the corresponding channel matrix is written as below:

$$H = \sqrt{\frac{N_t N_r}{N_{cl} N_{ray}}} \sum_{i=1}^{N_{cl}} \sum_{k=1}^{N_{ray}} \alpha_{ik}\, a_r(\phi_{ik}^r, \theta_{ik}^r)\, a_t(\phi_{ik}^t, \theta_{ik}^t) \qquad (5)$$

where $a_t$ and $a_r$ are the normalized antenna array response vector for transmitter and receiver respectively. $\phi_{ik}^t$ ($\theta_{ik}^t$) and $\phi_{ik}^r$ ($\theta_{ik}^r$) are the elevation angles of arrival and departure of the kth ray in the $i^{th}$ scattering cluster on the receiving and transmitting sides, respectively. $N_{cl}$ and $N_{ray}$ are number of scattering clusters and number of propagation paths in a cluster respectively. $\alpha_{ik}$ is the gain of the $k^{th}$ ray in the $i^{th}$ scattering cluster. For the uniform plane array (UPA) in the yz plane with L and B elements on the y and z axis respectively, the expression for array response vector is given by

$$a(\Phi, \Theta) = \frac{1}{\sqrt{L \times B}}\left[1, \ldots, e^{j(2\pi/\lambda)d(m\sin\theta\sin\phi + N\cos\theta)}, \ldots, e^{j\left(\frac{2\pi}{\lambda}\right)d((L-1)\sin\phi\sin\theta + (B-1)\cos\theta)}\right]^T \qquad (6)$$

where, LXB is the antenna array size, λ is the wavelength of the carrier signal, d is the spacing between the adjacent antenna elements, $0 \leq m < L$ and $0 \leq n < B$ are the y and z indices of an antenna element respectively. The purpose of the hybrid precoding is to maximize Spectral efficiency [11] of the system which is expressed as below:

$$S.E. = log_2 \left( \left| I_{N_s} + \frac{\rho}{N_s} R_n^{-1} W_A^H W_D^H H F_D F_A F_D^H F_A^H H^H W_A W_D \right| \right) \tag{7}$$

where $R_n = \sigma_N^2 W_A^H W_D^H W_A W_D$ represents the noise covariance matrix after combining.

2.3 Problem Formulation

The most common optimization problem is to maximize the overall spectral efficiency under the constraints. The optimization problem can be written as:

$$\max_{\{F_D, F_A, W_A, W_D\}} SE(F_D, F_A, W_A, W_D) \tag{8}$$

$$s.t. \|F_D F_A\|_F^2 \leq P$$
$$|F_A(i,j)|^2 = 1, \forall i, j$$
$$|W_A(i,j)|^2 = 1, \forall i, j$$

The first constraint is the total power constraint or the total power budget at the base station. The second and the third constrains are the constant modulus constraints for the analog precoder and combiner respectively. The optimization problem is non-convex due to the non-convexity of the constraints. The optimization problem is non-convex due to the non-convexity of the constraints on FA and WA. The objective function is the expression for the spectral efficiency. Finding the optimal solution for the optimization problem is challenging using the traditional methods over the given parameters FD, FA, WA and WD. For the solution, deep neural network (DNN) based approaches are being proposed.

## 3. Deep learning-based hybrid beamforming architecture

The input layer, the hidden layers and the output layer are the different units of the deep neural networks. All the nodes are connected each other where each connection has a specific weight. Figure 3 represents the fully connected hybrid precoding structure and the phase shifter configuration has comparable topology with neural networks. Also, the precoding processing in base band is similar to the mathematical model of neural network [19], [21]- [23]. Base band precoding process includes the multiplication between the input data sequence and the corresponding weight vectors. More than two weights are to be considered while mapping the hybrid precoding configuration to a multilayer neural network. The hidden layers receive the inputs from the input layer and provide the output with the help of activation function. The activation function decides where or not to activate a particular node. The output of the hidden layer is forwarded to the output layer [24] [25].

The signal processing in a digital beamformer is written as $x^D = F_D S$ where S and $x^D$ are the complex signals which include both phase and amplitude. S to $x^D$ transformation is the mapping process. So, in general it can be written as: $x^D = f(S)$. If ϕ is parameter that is trainable for a neural network, the mapping (transformation) operation can be written as:

$$f_\phi(s) = f_a(W.S + B). \tag{9}$$

where $f_a$ is activation function, W is weight matrix and B is bias of the neural network. The mapping operation can be viewed in the figure 4. Since deep neural network consists of multiple layers, the mapping operation can be rewritten as:

$$f(\phi, W) = f_\phi(.....f_\phi(f_\phi(s))). \tag{10}$$

The figure 5 shows the proposed DNN structure. Here, the dimension determines each training sequence length in the input layer. All the layers considered here are the fully connected layers. The input layer having 100 units is responsible for getting the input data features. To enable the encoding operation, two fully connected hidden layers implemented. The first two hidden layers include 300 units and 128 units respectively. After two encoding hidden layers, another hidden layer is introduced to add the noise. Before the output layer, two hidden layers having 100 and 64 units are added for the decoding purpose.

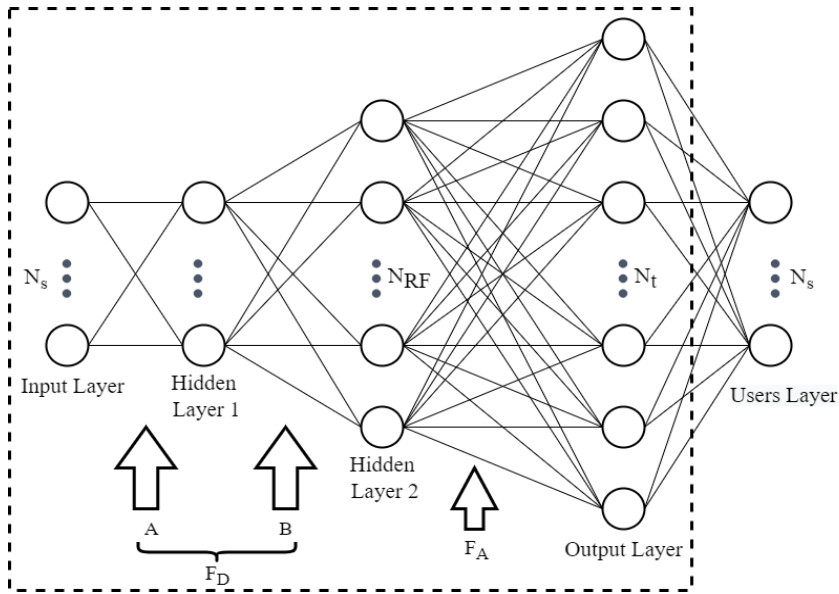

**Figure 4:** The Mapping of a Multi-Layer Neural Network

The ReLU function is used as an activation function for the input and the hidden layers and sigmoid function is used in the output layer because output layer has multiples outputs. The ReLU function is represented as:

ReLU(x) = max(x,0) $\forall$x and Sigmoid function is given by $f(x) = \frac{1}{1+e^{-x}}$. Now, in encoder stage of auto encoder, the encoder receives the input S and maps it to the f(s) as given in equation (10). In the decoder stage, the auto encoder maps f(s) to the reconstructed value $\hat{S}$ having the same dimension(shape) of S.

$$\hat{S} = f'_a(W' f_\phi(s) + B'). \tag{11}$$

where $f_a$ is the activation function in decoder part which may be some variation of f(s), W' and B' are also some variations in the initial value of W and B respectively. To obtain the best values of weights and biases, training is carried out using gradient descent algorithm [27]. Reducing the loss directly increases the average spectral efficiency. For the realization of the hybrid precoding

through training, the loss function is given by

$$\|F_{opt} - F_H\|_F = \sqrt{tr(F_{opt} - F_H)(F_{opt} - F_H)^H} \tag{12}$$

where $F_H = F_D F_A$ is the hybrid precoder, $\|.\|_F$ represents the Frobenius norm, $F_D$ and $F_A$ are digital and analog precoder respectively.

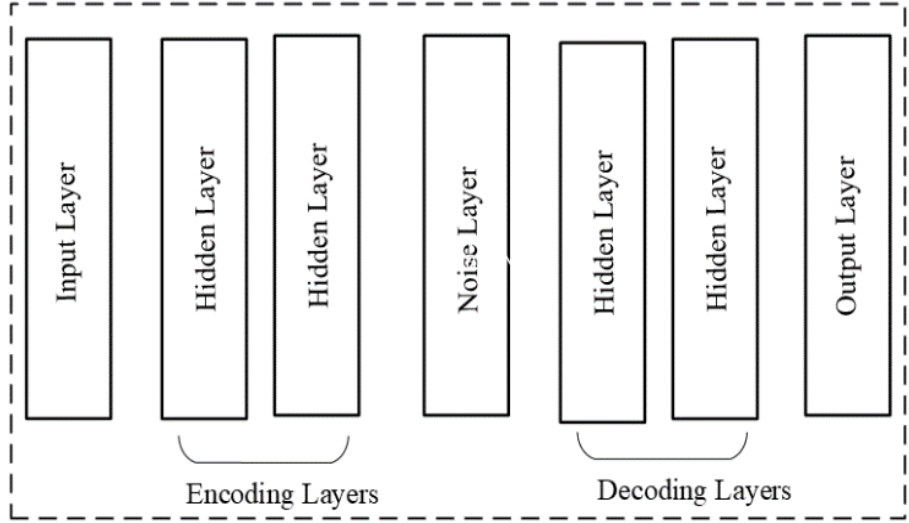

**Figure 5:** The DNN architecture for massive MIMO hybrid beamforming

The $F_{opt}$ is obtained from the singular value decomposition of channel matrix. For the training purpose, an auto encoder neural network is employed and network parameters are obtained. The auto encoder extracts the necessary features continuously during the propagation phase and filters out the unwanted information. It implements unsupervised learning algorithm which uses back propagation method. In auto encoder model, the stage includes the feature learning using unsupervised learning method. For every input, forward propagation takes place to get the output value. The mean square error (MSE) is implemented to calculate the deviation between the input and the output. Then back propagation of the error takes place through the network and the weights are updated. The auto encoder is constructed as:

$$F_{opt} = f(F_D F_A; N) \tag{13}$$

where N denotes sample data set and f(.) represents mapping process [17]. At first, $F_D$ and $F_A$ are initialized as empty matrices. The angle of arrival (AoA) and the angle of departure (AoD) are generated in random fashion. When AoA and AoD measurement error are increased, the SNR performance decreases due to this, the optimized precoding weights cannot be optimal values for the optimal performance. The bias between the $F_{opt}$ and $F_D F_A$ can be found corresponding to the output layer going through the numerous iterations. For the learning process, the data is divided into training data set, validation data set and testing data set. The training data set is used to know the weights between the nodes while validation data set help for fine tuning the performance. The test data set are applied to find the accuracy of the deep neural network. The optimizer is used to facilitate the training process by optimizing the weights.

## 4. Results and Discussion

The proposed DNN model is implemented using Keras and TensorFlow library in python. For data set, channel matrix (H) is generated according to the equation 5. The transmit power = 12 dBm, epochs = 100, batch Size = 250, learning rate =0.0001 and uniform planar array (UPA) is considered for both transmitter and the receiver. The number of RF chain at both transmitter and receiver is taken as 4. In the propagation environment modeling, for every channel matrix(H), $N_{CL}$=5 and $N_{ray}$= 5 for every cluster is provided. The random distribution of azimuth and elevation angles is taken for [0, 2π]. The frequency of operation is 28 GHz, bandwidth is 100 MHz and antenna spacing is half the wavelength. The normalized channel matrix is provided as input to the neural network. The data set is divided into training data set 80%, validation data set 10% and test data set 10%. The Adam optimizer is used for updating the weights of the neural network. The figures obtained from the simulation are shown in figure 6 to figure 13.

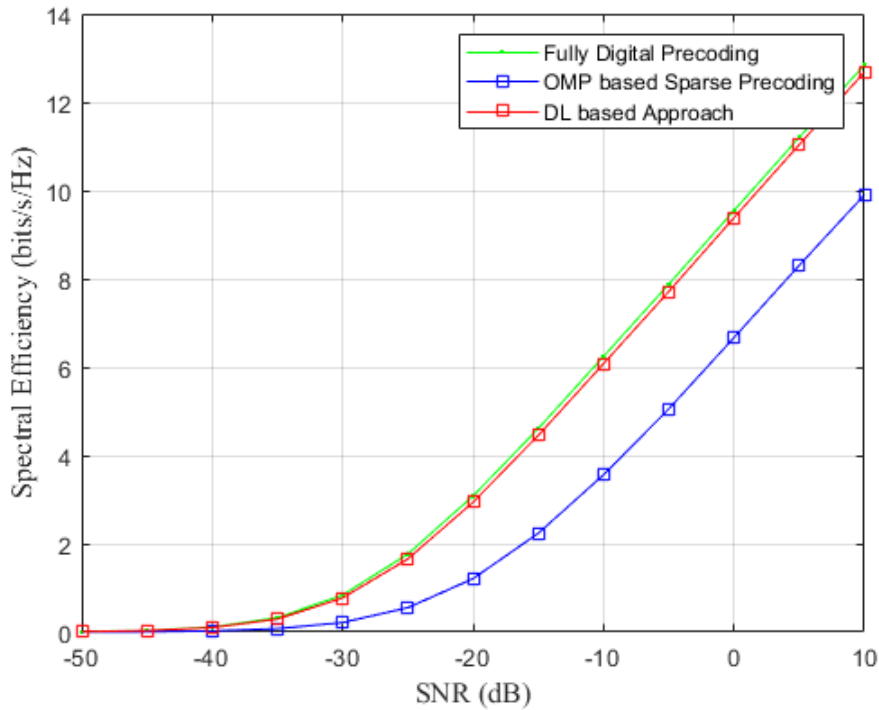

**Figure 6:** Spectral efficiency comparison in terms of SNR

Performance of the various hybrid precoding scheme is analyzed in terms of spectral efficiency, complexity, bit error rate and mean square error. In figure 6, the spectral efficiency is plotted as a function of varying SNR value and it illustrates the achievable spectral efficiency in a 64 x 16 massive MIMO system with uniform planar array at both transmitter and receiver. It is shown that the performance of the proposed approach based on deep learning is nearly equivalent to the optimal performance and the performance in terms of spectral efficiency is crucially superior than that of orthogonal matching pursuit (OMP) based sparse precoding [2].

The figure 7 compares the spectral efficiency for zero forcing hybrid precoding [6], OMP based sparse precoding [2] and proposed DNN based precoding. The result presented in the figure 7 depicts that there is significant performance gap among those three schemes and deep learning-based method has the best performance in terms of spectral efficiency observing for increasing number of the base station antennas. For zero forcing method, the spectral efficiency is lower because of its incapability of dealing with multipath environment. Also, it can be noted that deep learning-based scheme can provide approximately 6% to 10% higher spectral efficiency compared to traditional OMP based hybrid precoding. The figure 8 illustrates that when the number of base station antennas increases, the signal processing complexity also increases. The complexity is studied in terms of number of complex multiplications. From the figure 8 as an example, the number of complex multiplications for 256 antennas considering the OMP based precoding is approximately $3x10^6$ while that value is around $0.5x10^6$ for proposed DNN based method. So, the signal processing complexity can be reduced substantially by applying deep learning process for millimeter wave massive MIMO hybrid precoding optimization. If signal processing complexity decreases beams are generated faster and latency of the network also decreases. Hence, reducing the computational complexity, latency of the system can be decreased significantly by incorporating deep learning-based hybrid precoding.

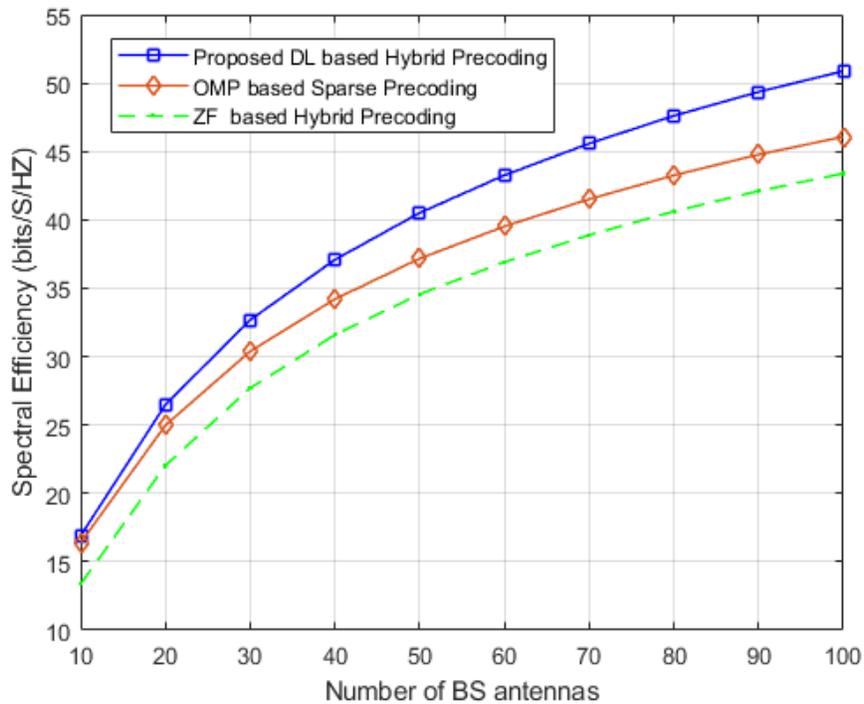

**Figure 7:** The spectral efficiency comparison among the proposed deep learning-based approach, OMP based sparse precoding and zero forcing based hybrid precoding

To analyze the robustness of the proposed deep learning-based hybrid precoding scheme, MSE versus number of iterations is plotted as shown in the figure 9. With the increasing number of iterations, the mean square error (MSE) goes on decreasing for both precoding strategies but becomes stable after around 15 iterations. The mean square error is lower for the proposed deep learning-based approach as compared to the OMP based traditional hybrid precoding technique. In figure 10, the transmit beam pattern is plotted using optimal weights and the hybrid beamforming weights. For this figure, 64 x 16 transmit antenna array with a directed beam width of 55 degrees in azimuth and 20 degrees in elevation is considered. The angular spread is taken as

10. The figure 10 presents that there are few numbers of dominant directions despite of the multipath environment.

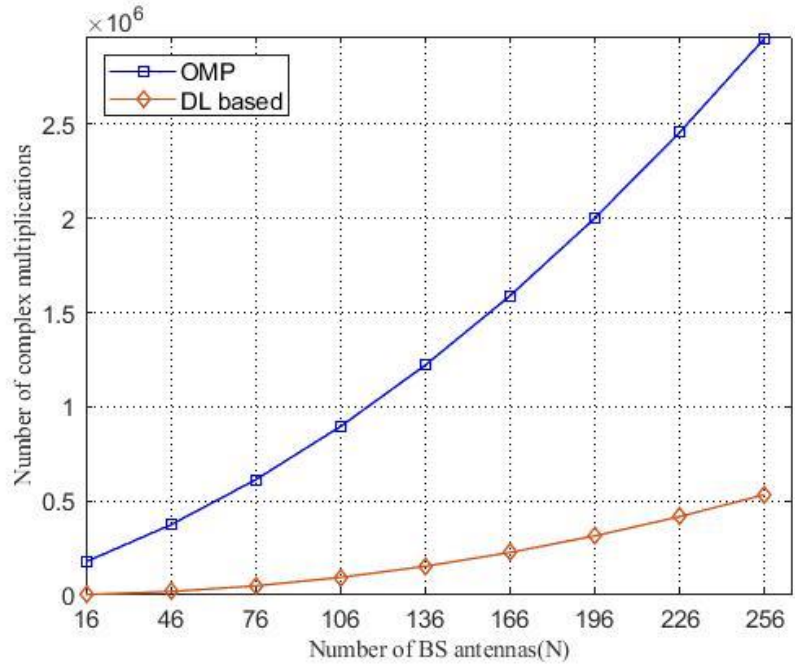

**Figure 8:** The complexity comparison between the proposed deep learning-based approach and OMP based sparse precoding

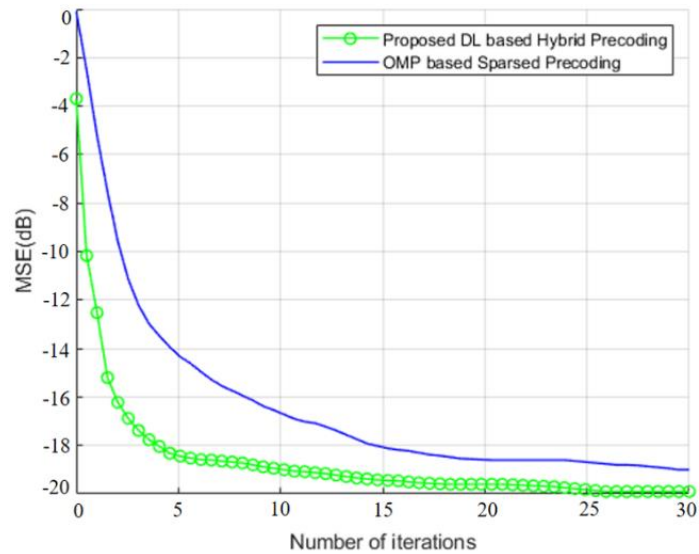

**Figure 9:** MSE versus number of iterations graph for the proposed deep learning-based hybrid precoding and the OMP based sparse precoding

The obtained transmit beam pattern implementing hybrid weight is similar with the beam pattern using optimal beamforming weights particularly for the beams in dominant direction if figure 10 is compared with the figure 11. In figure 12, the bit error rate (BER) performance is compared among the conventional OMP based approach, zero forcing scheme and the auto-encoder neural network assisted technique for solving hybrid beamforming optimization problem. Here, for the BER performance comparison, 16 QAM modulation is used.

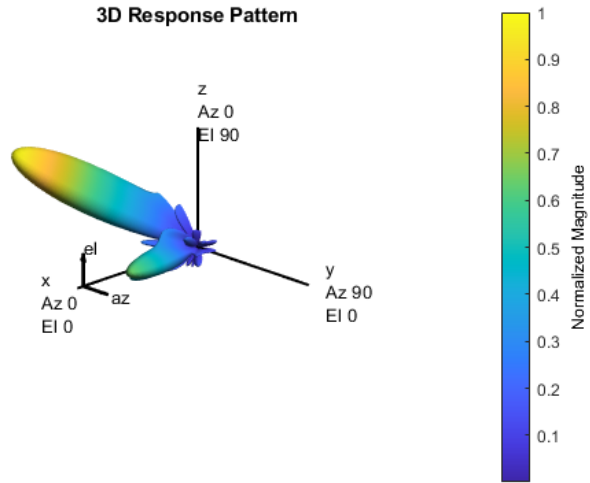

**Figure 10:** Transmit beam pattern using optimal weights

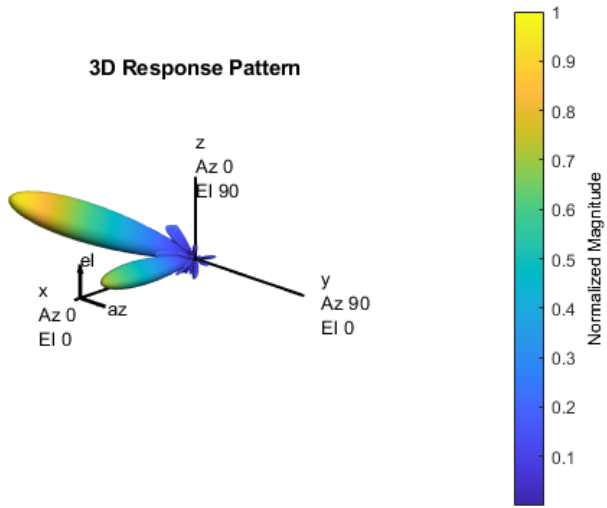

**Figure 11:** Transmit beam pattern using hybrid precoding weights

In this graph, the SNR gain achieved at any particular value of BER is considerable. Learning rate equal to 0.0001, the transmit antennas ($N_t$) = 64, the received antennas ($N_r$) = 16, number of RF chains at both transmit and receive antennas, ($N_{RF}$) = 5 are the parameters used for simulation. Bit error rate plays key role for the analysis of the link reliability. Among all the schemes, hybrid beamforming method outperforms and shows its superiority in terms of BER as shown in the figure 12. The graph as shown in figure 13 is plotted to check the model accuracy. Model is compiled before training and the testing. It can be remarked that the proposed model is working properly providing approximately 87% training accuracy and about 85% test accuracy.

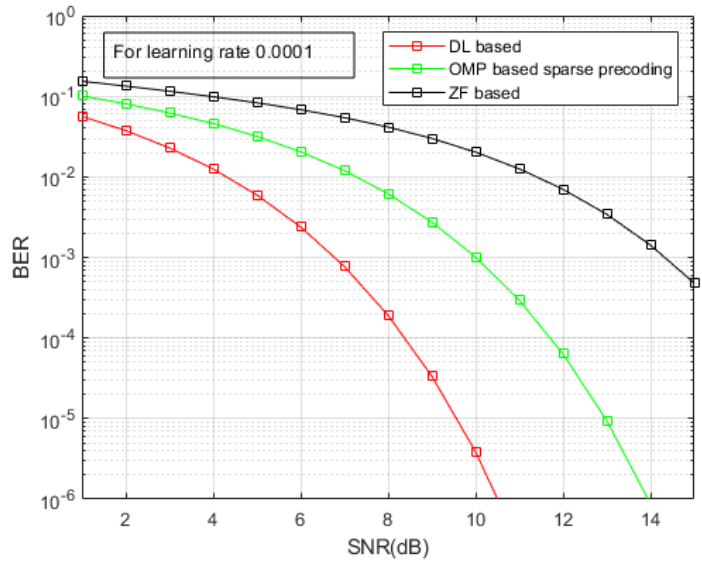
**Figure 12:** BER performance comparison

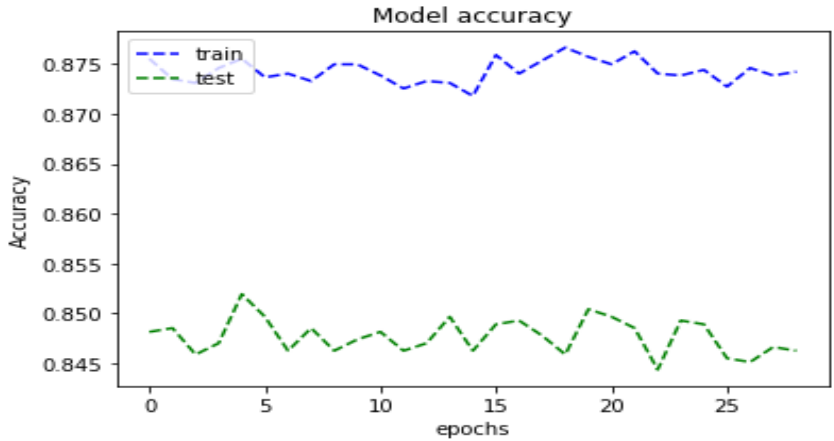
**Figure 13:** Graph for the model accuracy

## 5. Conclusion
The deep learning driven approach and other traditional methods are studied for solving the non-convex hybrid precoding optimization problem in millimeter wave massive MIMO system. The deep learning assisted scheme shows significant performance gap compared to the conventional methods in terms of spectral efficiency, complexity and bit error rate. The simulated results show that the deep learning-based approach outperforms the traditional hybrid precoding optimization procedures. The proposed deep learning implemented strategy provides the near optimal solution for non-convex hybrid precoding optimization problem. The results emphasize the requirement and importance of the hybrid precoding based on the deep learning for the millimeter wave massive MIMO system. For further study, reinforcement and supervised learning can be considered for the comparative analysis along with the different antenna array configurations and other channel parameters.